\documentclass[twocolumn]{aastex7}

\newcommand\cd{d$^{-1}$\,}

\usepackage[normalem]{ulem}
\submitjournal{ApJ}
\begin{document}

\title{\textit{Kepler} Discovery of GW Vir Pulsations of the Central Star of Planetary Nebula Kn 61}

\correspondingauthor{Paulina Sowicka}

\author[orcid=0000-0002-6605-0268,sname='Sowicka']{Paulina Sowicka}
\affiliation{Instituto de Astrofísica de Canarias, E-38205 La Laguna, Tenerife, Spain}
\affiliation{Universidad de La Laguna, Dpto. Astrofísica, E-38206 La Laguna, Tenerife, Spain}
\affiliation{Nicolaus Copernicus Astronomical Center, Polish Academy of Sciences, ul. Bartycka 18, PL-00-716, Warszawa, Poland}
\email[show]{paulina.sowicka@iac.es}  

\author[orcid=0000-0001-7756-1568,gname=Gerald, sname='Handler']{Gerald Handler} 
\affiliation{Nicolaus Copernicus Astronomical Center, Polish Academy of Sciences, ul. Bartycka 18, PL-00-716, Warszawa, Poland}
\email{gerald@camk.edu.pl}

\author[orcid=0000-0001-5941-2286,gname=J. J.,sname=Hermes]{J. J. Hermes}
\affiliation{Department of Astronomy \& Institute for Astrophysical Research, Boston University, 725 Commonwealth Avenue, Boston, MA 02215, USA}
\email{aaa@aaa.aa}

\author[orcid=0000-0002-0656-032X,gname=Keaton,sname=Bell]{Keaton J.\ Bell}
\affiliation{Department of Physics, Queens College, City University of New York, Flushing, NY-11367, USA}
\email{Keaton.Bell@qc.cuny.edu}

\begin{abstract}

We report the discovery of pulsations in the N-rich PG~1159-type central star of the planetary nebula Kn~61 based on one month of \textit{Kepler} Short Cadence observations. We detect four significant peaks in the frequency range consistent with $g$-modes excited in GW Vir stars. From the detected modes, we identify a mean period spacing of $\Delta\Pi=21.526(6)$~s for a sequence of three $\ell=1$ modes. This allows us to derive the asteroseismic mass of the star, which we estimate to be $0.551(6)~\mathrm{M}_{\odot}$, consistent with the one derived from the evolutionary tracks. We also characterize sporadic brightening events in the Long Cadence \textit{Kepler} light curve of Kn 61. If we assume these are caused by increases in effective temperature, we estimate their energies to be $\sim10^{40}$\,erg, though this may not be accurate as the mechanism for releasing so much energy is still unknown. 

\end{abstract}



\section{Introduction}
\label{sec:intro}

PG 1159 stars represent a transitional phase in stellar evolution between the post-asymptotic giant branch (post-AGB) phase and the white dwarf (WD) cooling stage. The post-AGB phase is arguably one of the least understood phases of the evolution of low- and intermediate-mass stars, making those stars important to study from an evolutionary point of view. These hydrogen-deficient pre-WDs are characterized by very high effective temperatures (75,000--200,000 K) and surface compositions rich in helium, carbon, and oxygen \citep{2006PASP..118..183W}. Many PG 1159 stars are observed as central stars of planetary nebulae (CSPNe), making them important probes for both stellar evolution and the late stages of nebular dynamics.

Notably, a subset of PG 1159 stars, known as GW Vir stars\footnote{The GW Vir family of pulsating pre-WD stars not only consists of PG 1159 stars, but also of hybrid-PG 1159 stars, [WC] stars, and [WC]-PG 1159 transition objects.}, display non-radial $g$-mode pulsations, characterized by low amplitudes (1 mmag--0.15 mag) and short periods \citep[300--6000 s, see, e.g.,][]{2019A&ARv..27....7C}, driven by the $\kappa-\gamma$-mechanism operating in the partial ionization zones of carbon and oxygen \citep{1983ApJ...268L..27S,1984ApJ...281..800S}. As with other groups of pulsating stars, asteroseismic studies of these pulsations provide invaluable insights into the internal structures and evolutionary processes of these stars, with the group prototype, PG 1159-035 ($\equiv$ GW Vir), being one of the best studied stars of this type \citep[and references therein]{2022ApJ...936..187O}.
The effectiveness of asteroseismic methods relies critically on both the number of detected pulsation modes and the depth within the stellar interior that these modes can probe — the broader the range, the more information on the stellar interior that can be retrieved. Consequently, stars exhibiting rich pulsation spectra are the most favorable targets for detailed asteroseismic studies.

Recent observational campaigns have expanded our understanding of the GW Vir instability strip, which is unique among WD instability strips as it is not pure, meaning that not \textit{all} stars within its borders show pulsations \citep[see, e.g.,][]{2008PASP..120.1043F}. A current hypothesis, based on combined photometric
and spectroscopic observations, states that there is a separation within PG 1159 stars: all N-rich (about 1\% atmospheric N/He abundance) PG 1159 stars are pulsators, while N-poor ones (below about 0.01\% N/He) were identified as non-pulsators - the so-called nitrogen dichotomy. This apparent dichotomy seems to hint at a potential link between a very late thermal pulse (sometimes referred to as a born-again event) and pulsations (\citealt{1998A&A...334..618D}; \citealt{2021ApJ...918L...1S}). 
\citet{2023ApJS..269...32S} conducted time-series photometry of 29 PG 1159 stars, leading to the discovery of pulsations in the central star of planetary nebula Abell 72, and derived the most robust pulsator fraction to-date for PG 1159 stars within the GW Vir instability strip: 36\% (24 out of 67 stars). Interestingly, they also found a few objects that appear to be pulsators but show no evidence of being N-rich, further challenging the current hypothesis. 
These findings underscore the significance of continued searches for pulsating PG 1159 stars to refine the boundaries of the instability strip and enhance asteroseismic models, together with high-quality spectroscopy to determine $T_{\mathrm{eff}}$ and $\log{g}$, which are needed to establish the stars' positions within the instability strip.

We present the discovery and frequency analysis of the detected peaks of multi-periodic pulsations in the central star of the planetary nebula Kn 61, spectroscopically classified as a PG 1159-type star, based on \textit{Kepler} Short Cadence observations from Quarter 12.3. This discovery contributes to the growing catalog of GW Vir pulsators and offers a new target for detailed asteroseismic analysis. We also characterize sporadic brightening events that have been previously noted \citep{2015MNRAS.448.3587D} in the Long Cadence \textit{Kepler} data from Quarters 10-13. 

\section{The central star of planetary nebula Kn 61}

\begin{figure}
    \centering
    \includegraphics[width=\columnwidth]{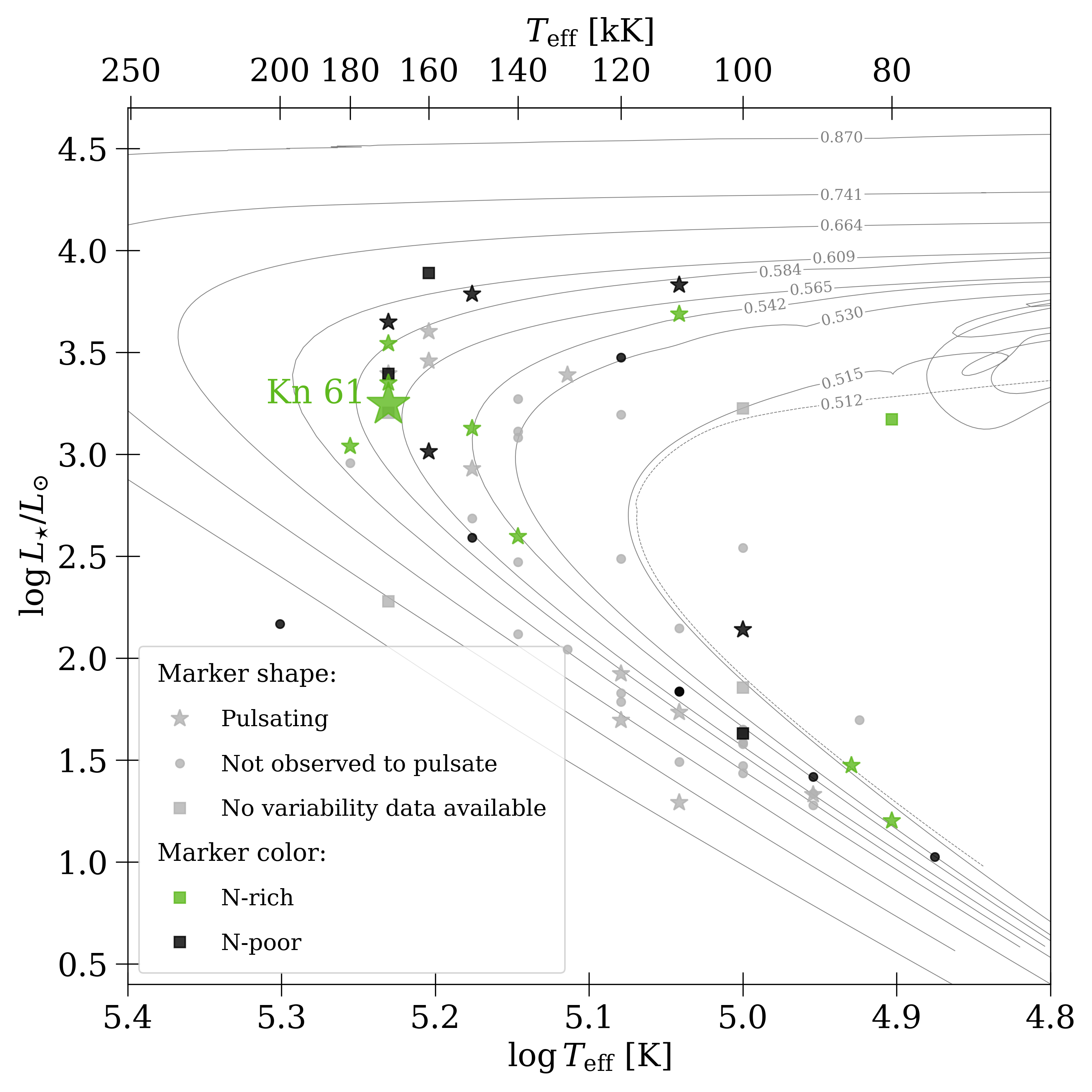}
    \caption{Position of Kn 61 and other PG 1159 stars in the theoretical H-R diagram. The sample (including luminosities and other information) comes from Table 2 in \citet{2023ApJS..269...32S}. Star symbols: pulsating PG 1159 stars; circles: nonvariable; squares: with no reported photometric observations. N-rich PG 1159 stars are shown with green symbols, while N-poor ones are shown with black symbols, and stars without N abundance reported in the literature are shown with gray symbols. Lines represent evolutionary tracks from \citet{2006A&A...454..845M}: solid lines—VLTP (from left to right, ﬁnal masses: 0.870, 0.741, 0.664, 0.609, 0.584, 0.565, 0.542, 0.530, 0.515 M$_{\odot}$); single dashed line—LTP (0.512 M$_{\odot}$).}
    \label{fig:Kn61}
\end{figure}

The planetary nebula Kn 61 (also known as the Soccer Ball Nebula) was discovered in a deep sky survey to identify new planetary nebulae \citep{2012IAUS..283..414K}. The first spectra of the central star (R.A.: 19h 21m 38.9356s, Dec: $+38\arcdeg\ 18\arcmin\ 57.217\arcsec$ (J2000), Gaia G = 18.25 mag, Gaia BP-RP=-0.36, $\varpi_{\mathrm{Gaia}}=0.14 \pm 0.11$ mas, \textit{Kepler} target KIC 3231337, hereafter referred to as Kn 61 for simplicity) suggested it is of hydrogen-deficient PG~1159-type with $T_{\mathrm{eff}}=120{,}000$~K and $\log{g}=7.0$~cm\,s$^{-2}$ \citep{2014AJ....148...57G}. New spectra obtained by \citet{2023MNRAS.521..668B} confirmed the spectral classification and allowed a preliminary re-determination of atmospheric parameters: $T_{\mathrm{eff}}$ possibly close to $170{,}000$~K and $\log{g}\simeq6.5$~cm\,s$^{-2}$, based on the wings of the \ion{He}{2} absorption lines. The presence of \ion{N}{5} lines indicated the ``N-rich'' variety of PG~1159-type stars, which are usually pulsators of the GW Vir type \citep[e.g.,][]{2021ApJ...918L...1S}.
In Figure~\ref{fig:Kn61} we show the location of Kn 61 in the Hertzsprung-Russell (H-R) diagram with other known PG 1159 stars, placing it among the confirmed N-rich pulsating PG 1159 stars. The luminosities of the objects shown in Fig.~\ref{fig:Kn61} were derived by \citet{2023ApJS..269...32S} from distances based on Gaia DR3 parallaxes \citep{2021AJ....161..147B} combined with bolometric corrections.

Ground-based photometric observations of the central star were first reported by \citet{2014AJ....148...57G}. On the basis of observations over 16 consecutive nights, each $2-3$ hr long, the authors did not detect variability on such a short time scale (with a scatter of the order of 0.028 mag), or on timescales typical for pulsations. However, the nightly averaged data revealed a statistically significant variation with a frequency of 0.176~\cd (a period of $5.7\pm0.4$~d) and semi-amplitude $A\approx 0.02$~mag.   

Kn 61 lies in the Field of View (FOV) of the \textit{Kepler} mission \citep{2010ApJ...713L..79K}, and was observed in Quarters (Q) $10-13$ in Long Cadence (LC, 30 min; 2011~Jun~28 -- 2012~Jun~27\footnote{Exact dates for each quarter: Q10: 2011 Jun 28 -- 2011 Sep 27, Q11: 2011 Sep 29 -- 2012 Jan 04, Q12: 2012 Jan 05 -- 2012 Mar 28, Q13: 2012 Mar 29 -- 2012 Jun 27}), and in Q12.3 in Short Cadence (SC, 1 min; 2012~Mar~01--28) modes. The \textit{Kepler} data in LC and SC span three months and a month, respectively. \citet{2015MNRAS.448.3587D} analyzed four quarters of \textit{Kepler} LC data, showing that the light curve is characterized by stochastically recurring brightenings that recur roughly every 2-12~d, each with a duration of several hours up to 2~d. These brightening events cause an increase in brightness of $\approx 80-140$~mmag, and have oddly triangular pulse shapes (see Figure 9 from \citealt{2015MNRAS.448.3587D}). The authors were not able to unambiguously determine the origin of this variability, ruling out irradiation, ellipsoidal variability, or eclipses, with alternative scenarios involving binarity, wind, and accretion. The light curves resemble the outbursts in pulsating DA white dwarfs (e.g., \citealt{2015ApJ...809...14B,2015ApJ...810L...5H,2016ApJ...829...82B}), as discussed in more detail by \citet{2023MNRAS.521..668B}. We note that the first PG~1159 star with reported outbursts (and also pulsating) was the central star of planetary nebula Longmore 4 (Lo 4, \citealt{1990AJ....100..788B, 1992A&A...259L..69W, 2014AJ....148...44B}), however the reported outbursts were only spectroscopic, with no photometric outbursts reported for this star to date (for more details, see Section~\ref{sec:outbursts}).

\section{Observations and Frequency Analysis}

We downloaded and analyzed the \textit{Kepler} SC data processed by the \textit{Kepler} Asteroseismic Science Operations Center (KASOC,\footnote{\url{https://kasoc.phys.au.dk}} using version~4 of the KASOC-corrected light curves). KASOC ran a filter to remove transits, long-term trends, and bad data points from \textit{Kepler} timeseries \citep{KASOCfilter2011}. Given the \textit{Kepler} pixel scale of $\sim$4\arcsec\ pix$^{-1}$, we examined the neighborhood of Kn~61 and found no bright nearby stars that would significantly contaminate the photometry, which is consistent with the crowding metric, \texttt{CROWDSAP}\footnote{\texttt{CROWDSAP} gives the ratio of target flux to total flux in optimal aperture.} keyword value of 0.87. We converted the fluxes from e$^-$/s to \textit{Kepler} magnitudes using the reference flux for a 12th magnitude star, f12\footnote{\url{https://nexsci.caltech.edu/workshop/2012/keplergo/CalibrationSN.shtml}}. After removing outliers, the final light curve consisted of $32{,}939$ points.

\begin{figure*}
    \centering
    \includegraphics[width=\textwidth]{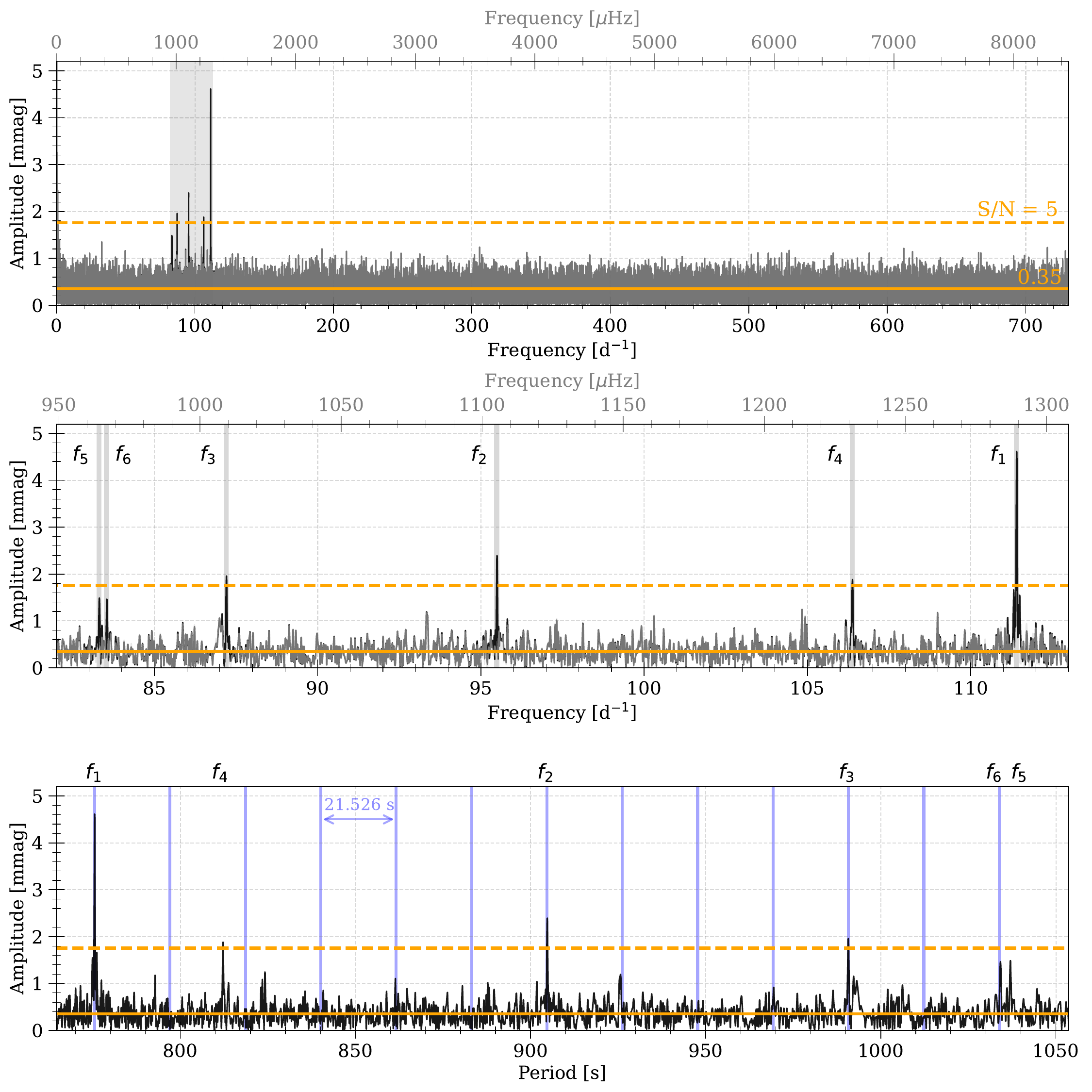}
    \caption{\textit{Top:} Fourier amplitude spectrum of \textit{Kepler} SC data up to the Nyquist frequency (black). Residuals after prewhitening 6 peaks are shown in gray. Orange lines show the mean noise level (solid) and the detection threshold of $S/N=5$ (dashed). The shaded region is shown in the middle panel. \textit{Middle:} A zoom-in view into the low frequency (shaded) region. Each accepted frequency is labeled according to the decreasing amplitude and indicated by a vertical gray line. Horizontal lines are the same as in the top panel. \textit{Bottom:} Fourier spectrum for the low-frequency region shown in the period domain. Blue vertical lines show the theoretical constant period spacing pattern of 21.526~s starting from $f_1$. Horizontal lines are the same as in the top panel.}
    \label{fig:FT}
\end{figure*}

We performed frequency analysis using \texttt{Period04} \citep{2005CoAst.146...53L}. In the Fourier amplitude spectrum, calculated up to the Nyquist frequency of 731.9 \cd, we detected four peaks satisfying the criterion of S/N$\geq$5 with a mean noise level of 0.35 mmag (following the results from \citealt{2015MNRAS.448L..16B}). We report the detection of two additional, formally insignificant peaks (S/N=4.1). All those peaks satisfy the Rayleigh resolution criterion of $\Delta f \geq 1.5/\Delta T = 0.055$~\cd \citep{1978Ap&SS..56..285L}. Frequencies, amplitudes, and phases were determined by simultaneously fitting a nonlinear least-squares solution to the data using \texttt{Period04}. The formal solution is presented in Table~\ref{tab:frequencies}. Figure~\ref{fig:FT} shows the Fourier amplitude spectrum up to the Nyquist frequency (upper panel), and a zoom-in into the $82-113$~\cd region (middle panel), where the peaks were detected. The peaks are labeled according to decreasing amplitude. We did not detect any frequency/amplitude/phase variations in the data set, as prewhitening all six peaks did not leave significant residuals.

\begin{deluxetable}{cDDccc}
    \tablewidth{\textwidth} 
\tablecaption{Formal frequency solution for Kn 61 
 \label{tab:frequencies}}
\tablehead{\colhead{ID} & \multicolumn2c{Freq.}& \multicolumn2c{Period} & \colhead{Amp.} & \colhead{S/N} & \colhead{$\ell$} \\
\colhead{} & \multicolumn2c{(d$^{-1}$)} & \multicolumn2c{(s)} & \colhead{(mmag)} & \colhead{} & \colhead{}
} 
\decimals
 \startdata
$f_1$ & $111.4087(13)$ & $775.5228(93)^*$ & $4.60(28)$ & $13.1$ & $1$ \\
$f_2$ & $95.4933(24)$ & $904.7760(48)^*$ & $2.40(28)$ & $6.9$ & $1$ \\
$f_3$ & $87.2077(30)$ & $990.7388(39)^*$ & $1.95(28)$ & $5.6$ & $1$ \\
$f_4$ & $106.3792(31)$ & $812.1891(37)$ & $1.86(28)$ & $5.3$ & \\ 
\multicolumn{8}{c}{\dotfill} \\ 
$f_5$ & $83.3167(40)$ & $1037.0067(29)$ & $1.43(28)$ & $4.1$ & \\
$f_6$ & $83.5423(40)$ & $1034.2070(29)$ & $1.43(28)$ & $4.1$ & \\
\enddata
\tablecomments{Frequencies below 2~d$^{-1}$ are omitted, due to long-term instrumental systematics. $f_5$ and $f_6$ are formally insignificant. Asterisks mark the periods we used for the fit shown in Figure~\ref{fig:period_spacing_fit}.
}
\end{deluxetable}

\subsection{Rotational splitting?}

To assess the asteroseismic potential of Kn 61, we first searched for rotationally split multiplets. For a spherically symmetric star, the frequencies of all nonradial modes with a given spherical degree $\ell$ and radial order $k$ are indistinguishable. Departures from the spherical symmetry, caused by various effects like, e.g., rotation or magnetic fields, can lift this degeneracy. In most stars, it is the (slow) rotation that is responsible for that lifting, splitting the modes into $2\ell + 1$ components. It is therefore expected to see three $(m=+1, m = 0, m = -1)$ components for $\ell = 1$ modes, five for $\ell = 2$ modes, and so on, if all components are observed. Modes with $\ell = 1$ or 2 are most likely to be observed due to geometric cancellation of higher degree modes \citep{1977AcA....27..203D}. Identifying rotational multiplets allows the identification of the $\ell$ and $m$ numbers of a mode, and determination of the mean frequency splitting, from which the rotation period of the star
can be inferred. 

The Fourier spectrum of Kn 61 is sparse, and we were only able to identify one potential doublet with a frequency separation of $f_6-f_5=83.5423-83.3167$~\cd$=0.2256$~\cd. If we assume that these are adjacent components of an $\ell=1$ triplet (i.e., $m=+1$ and 0 or 0 and $-1$), the observed separation corresponds directly to the rotational splitting. If instead they represent the outer components ($m=+1$ and $-1$), the true splitting would be half of the observed separation. The frequency splitting can be obtained from $\delta\nu_{lkm}=m\Omega_{rot,l}(1-C_{kl})$ under the assumption of slow and solid rotation, where $C_{kl}$ are Ledoux coefficients \citep{1958HDP....51..353L}. In the asymptotic limit of high radial order $g$ modes ($k\gg l$), the Ledoux coefficients adopt a simple form: $C_{l,k}\sim [\ell(\ell+1)]^{-1}$, which for a dipole $(\ell=1)$ mode are simply $C_{1,k}\sim0.5$. Depending on whether the observed doublet corresponds to adjacent or outer components of an $\ell=1$ triplet, the inferred rotation period of Kn 61 is $\approx 2.22$~d or $\approx 4.43$~d, respectively. In comparison with rotation periods of some other GW Vir stars obtained via asteroseismology (see, e.g., Table 10 in \citealt{2019A&ARv..27....7C}), the rotation period of over 53 hours for Kn 61 seems rather long. This might mean that either Kn 61 is a relatively slow rotator, or our identification of rotational splitting is not correct. 

\subsection{Mean period spacing}

Next, we searched for a mean period spacing. The asymptotic theory of nonradial oscillations gives a prescription for determination of a mean period spacing for a chemically homogeneous star: in the asymptotic limit $(k\gg\ell)$, consecutive (high) radial order ($k$) modes of the same spherical degree ($\ell$) are predicted to be approximately equally spaced in period (e.g., \citealt{1980ApJS...43..469T}), and the period spacing $\Delta\Pi_{\ell}$ is given by 
\begin{equation}
    \label{eq:pi0}
    \Delta\Pi_{\ell}=\frac{\Pi_{0}}{\sqrt{\ell(\ell+1)}}, 
\end{equation}
where $\Pi_0$ is a constant: $\Pi_0=2\pi^2\left( \int\frac{N}{r} dr \right)^{-1}$, where $N$ is the Brunt-V\"ais\"al\"a frequency. In the case of PG 1159 stars, which are chemically stratified, the observed period spacings show regular departures from a uniform pattern, as a result of resonant mode trapping by the composition gradient at the base of the He-rich surface layer \citep{1994ApJ...427..415K,2006A&A...454..863C}.

\begin{figure*}[t]
    \centering
    \includegraphics[width=1.0\textwidth,
        % height=0.72\textheight,
        % keepaspectratio
        ]{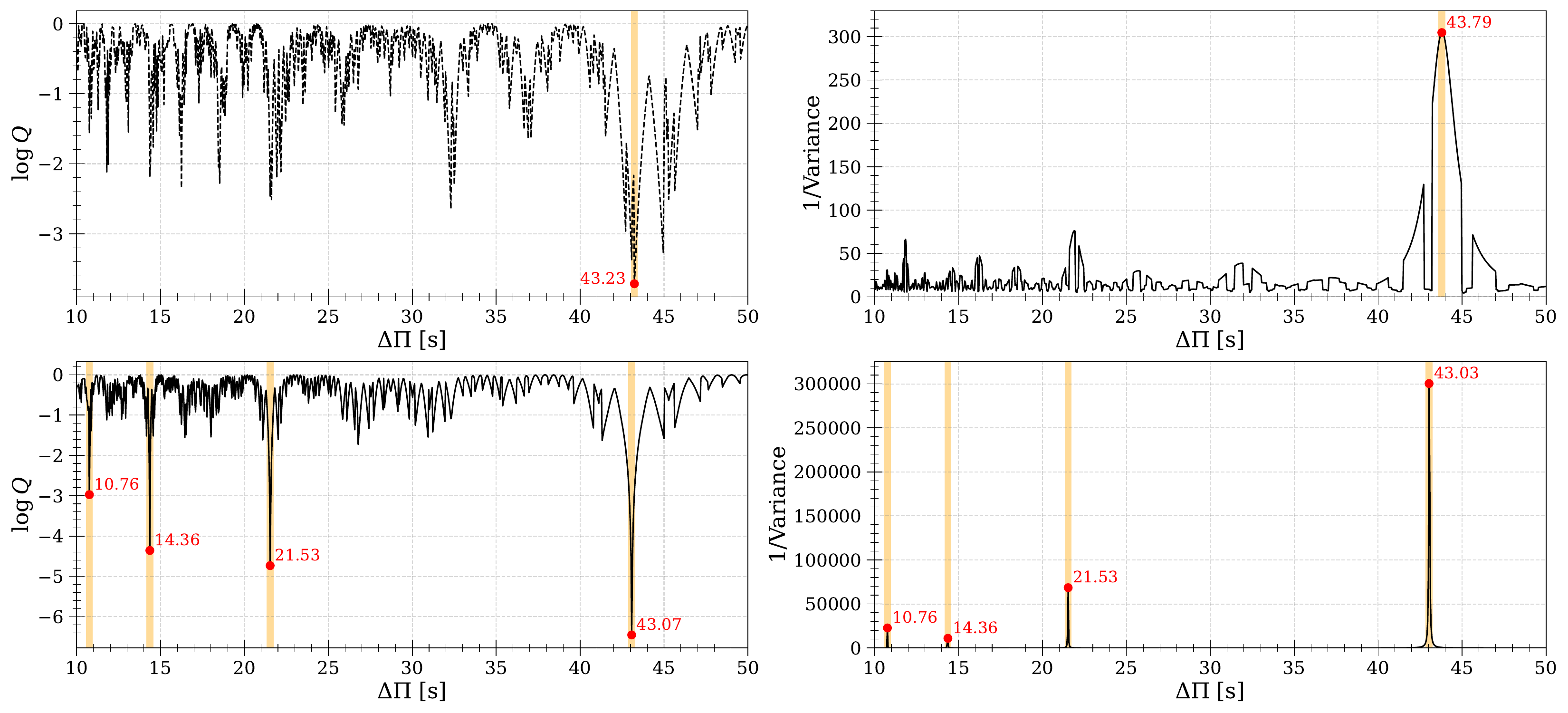}
    \caption{Results of the Kolmogorov-Smirnov statistical test (left) and Inverse-Variance method (right) for all detected periods from Table~\ref{tab:frequencies} (upper panels) and the subset of three modes marked with an asterisk in that Table (lower panels).}
    \label{fig:ks_test}
\end{figure*}

A closer inspection of the period differences calculated between the detected modes revealed a tentative spacing near 43 s. We employed standard techniques to search for period spacing values typical for GW Vir pulsators: the Fourier transform method \citep{1997MNRAS.286..303H}, the Inverse Variance method \citep{1994MNRAS.270..222O}, and the Kolmogorov-Smirnov \citep{1988IAUS..123..329K} statistical test. In the Fourier transform method we perform a Fourier analysis of the detected periods with arbitrarily assigned amplitudes of unity and look for statistically significant maxima. In this case we did not find any significant value indicating a mean period spacing, likely because of the sparse set of detected modes. In the Inverse Variance technique the spacing is chosen based on fitting equally spaced periods to the observed set of periods, and finding such a spacing that maximizes the inverse variance of the residuals of the model fit. In the Kolmogorov-Smirnov test, as used in a backwards fashion by \citet{1988IAUS..123..329K}, the uniform period spacings will appear as minima in Q (probability of a random distribution). Both those methods gave consistent results and pointed at the constant period spacing of about 43~s and its harmonics (approx. 21.5~s, 14~s, and 11~s), in the range of period spacings of 10--50~s.
A~43~s period spacing would be unusually large for a GW Vir star, pointing towards very small masses not supported by evolutionary models \citep{2006A&A...454..845M}. Therefore, we proceed assuming $\approx 21.5$~s as our trial period spacing value. In Figure~\ref{fig:ks_test} we show the results of the K-S and I-V tests in two cases -- using all modes from Table~\ref{tab:frequencies} or a subset of three modes marked with an asterisk. 

\begin{figure}
    \centering
    \includegraphics[width=\columnwidth]{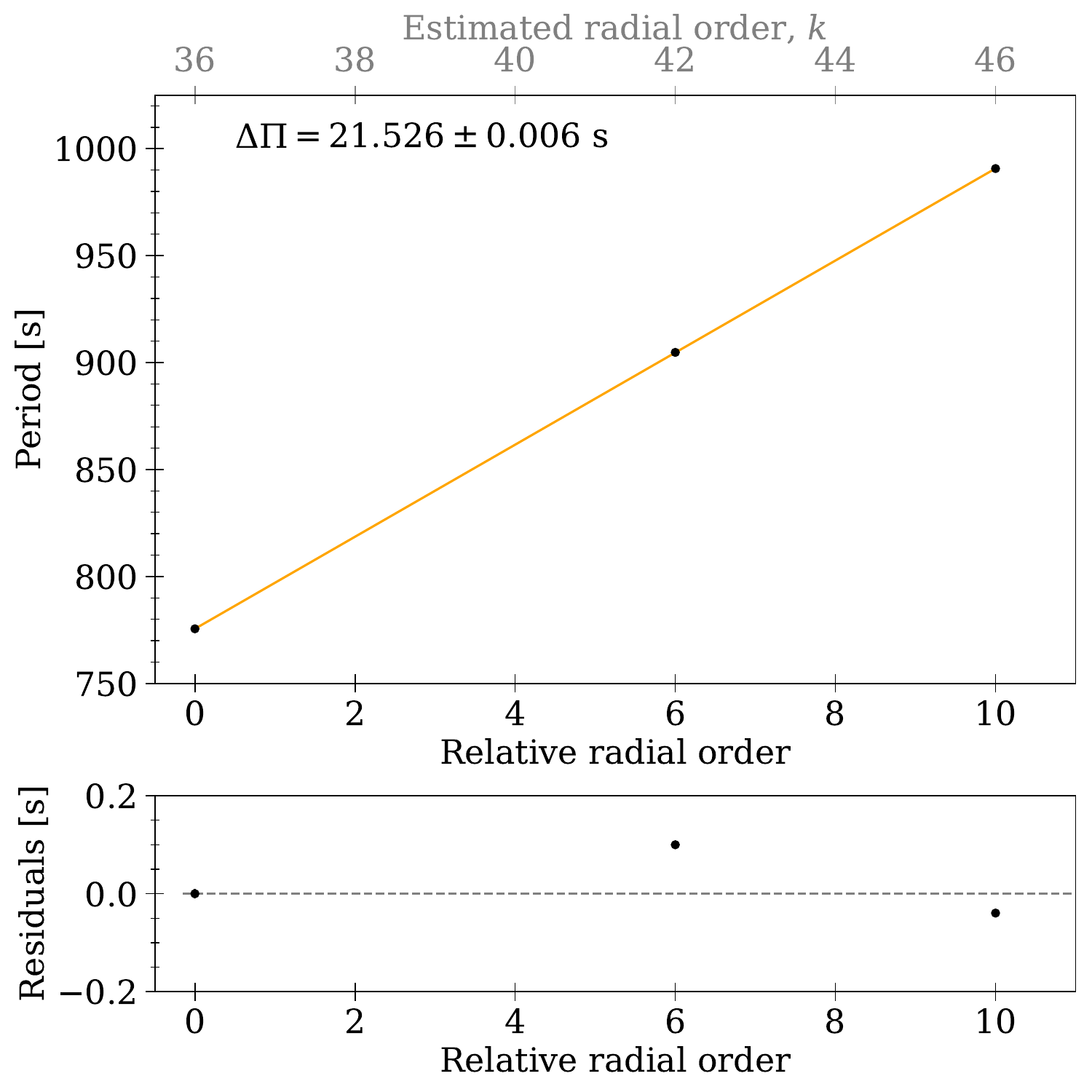}
    \caption{\textit{Top:} A linear least-squares fit to $f_{1}$, $f_{2}$, and $f_{3}$ modes, marked in Table~\ref{tab:frequencies}. \textit{Bottom:} Residuals after subtracting the fit from the observed periods. Note: Measured period uncertainties are plotted, but are smaller than the size of the points.}
    \label{fig:period_spacing_fit}
\end{figure}

To refine the period spacing, we used a weighted linear least-squares fit to the subset of three periods that we identify as $m=0$ modes, marked with an asterisk in Table~\ref{tab:frequencies}, and shown in the upper panel of Figure~\ref{fig:period_spacing_fit}. We obtained the period spacing of $\Delta\Pi_{\mathrm{fit}}=21.526(6)$~s. In Figure~\ref{fig:FT} (bottom panel) we show the Fourier amplitude spectrum in the period domain, and we indicate a theoretical sequence of constant period spacing of $21.526$~s obtained from the fit, propagated starting from $f_1$. It appears that $f_1$, $f_2$, and $f_3$ are part of a sequence of equally spaced modes, with several components of the sequence in between missing. We also note that, although $f_6$ is formally below the adopted detection threshold, its period is consistent with the spacing defined by $f_1$, $f_2$, and $f_3$ modes, and it aligns well with the sequence shown in the bottom panel of Figure~\ref{fig:FT}, providing additional support to the derived mean period spacing and increasing confidence in $f_6$. Possible explanations for the remaining two periods, $f_4$ and $f_5$, not aligning with the pattern in Figure~\ref{fig:FT} include: they could be $\ell=1$ modes affected by mode trapping, or $\ell=2$ modes (or higher), or some components of rotationally split modes, or they are caused by noise (especially $f_5$ which formally is below the detection threshold). Moreover, the spacings between $f_5$ or $f_6$ and the remaining modes may indicate that $f_6$ is the more likely $m=0$ component of the two.
In the bottom panel of Figure~\ref{fig:period_spacing_fit}, we show the residuals from subtracting the fit from the observed periods. The scatter of the periods around the best-fitting spacing is 0.07~s, far exceeding the formal period uncertainties, and could be attributed to departures from the asymptotic regime. 
The modes do not form a sequence of consecutive modes -- the missing modes in between might suggest that either mode trapping is strongly affecting some of those modes, they may not be excited in the star, or other cancellation/observational effects play a role.

For high-order $g$ modes, the asymptotic relation gives the periods as (see, e.g., \citealt{1994ApJ...427..415K})
\begin{equation}
    \label{eq:periods}
    P_{\ell, k} \simeq \frac{\Pi_{0}}{\sqrt{\ell(\ell+1)}} (k + \epsilon),
\end{equation}
where $\epsilon$ is a small number and represents a phase shift. To determine the absolute radial orders ($k$), we compared the observed periods to the asymptotic relation $P_{\ell, k}\approx\Delta\Pi_{\ell}(k+\epsilon)$. Using our determined period spacing of $21.526(6)$~s for $\ell=1$ modes, we tested a range of integer offsets to find the most physically consistent solution. By examining starting values in the range 30 to 45, we identified that an offset corresponding to radial orders $k \approx$ 36, 42, and 46 for the three detected modes ($f_{1}$, $f_{2}$, and $f_{3}$) minimizes the residual phase shift, with a mean value of $\epsilon \approx 0.03$. Given the small number of detected modes, this assignment is approximate and primarily intended to indicate the high-order nature of the pulsations. We added the estimated radial order axis in Figure~\ref{fig:period_spacing_fit}.

To place our radial order assignments in context, we compared the detected periods with those from asteroseismic modeling of the well-studied GW Vir star RX~J2117.1+3412, which is also a central star of a planetary nebula and has a similar effective temperature ($\sim$170,000~K) and a comparable location in the H-R diagram. In RX~J2117.1+3412, the shortest-period unstable $\ell=1$ mode occurs at $\sim757$~s with radial order $k=34$ \citep[Table 14]{2021A&A...645A.117C}, up to $\sim1190$~s and $k=53$. In Kn 61, our shortest-period detected mode at $\sim775$~s corresponds to $k\approx36$, with the next modes at $k\approx42$ and 46. The ranges of detected periods and corresponding radial orders are in very good agreement, demonstrating that our tentative $k$ assignments for Kn 61 are fully consistent with theoretical predictions for $\ell=1$ $g$-modes in a similar GW Vir star.

\subsection{Estimation of the asteroseismic mass}

Finding a constant period spacing in the Fourier spectrum of Kn 61 has important consequences: it allows the mode identification of the spherical degree ($\ell$), and based on that the determination of the asteroseismic mass. Based on asteroseismic calculations for GW Vir stars, a period spacing of $21.526(6)$~s is indicative of dipole ($\ell=1$) modes \citep{2006A&A...454..863C}, therefore we assigned that to the three identified modes in Table~\ref{tab:frequencies}. With the period spacing value assigned to the $\ell=1$ modes, we used Equation~\ref{eq:pi0} to calculate $\Pi_{0}=30.442(8)$~s, which we can directly compare with Figure 2 from \citet{1994ApJ...427..415K}. We also calculated $\Pi_{0}$ for alternative scenarios, that the true period spacing is about 43~s and either of those spacings is for $\ell=2$ modes, but neither alternative scenario matches the asymptotic period spacing sequences shown in the aforementioned Figure, strengthening our initial identification. We then used the approximate relation for $\Pi_{0}$ from \citet{1994ApJ...427..415K} to derive the asteroseismic mass:
\begin{equation}
    \Pi_{0} \cong 15.5 \left( \frac{M}{~\mathrm{M}_{\odot}}\right)^{-1.3} \left( \frac{L}{100 ~\mathrm{L}_{\odot}} \right)^{-0.035} \left( \frac{q_y}{10^{-3}} \right)^{-0.00012}. 
    \label{eq:mass}
\end{equation}
The dependence on the He fraction on the surface ($q_y$) is negligible for the approximate determination of the asteroseismic mass in our case and we omit this term. \citet{2023ApJS..269...32S} derived luminosities for the whole sample of PG 1159 stars, including Kn 61. Derivation of luminosities was based on the data from Gaia DR3 \citep{2016A&A...595A...1G,2023A&A...674A...1G}, including the geometric distance from \citet{2021AJ....161..147B} based on parallax but with a large error of 80\%. 
We therefore recomputed the luminosity of Kn 61 following the procedure described in \citet{2023ApJS..269...32S}, adopting the distance of $4.17\pm0.75$~kpc from \citet{2016MNRAS.455.1459F} and got $\log{L_{\star}}=3{.}24\pm0.17\ L_{\odot}$, a lower value than the one from \citet{2023ApJS..269...32S}. 
We used that value to obtain the asteroseismic mass of Kn 61 from Equation~\ref{eq:mass}: $m_{\mathrm{astero}}=0.551(6)~\mathrm{M}_{\odot}$. Using the luminosity and $\log{T_{\mathrm{eff}}}=5{.}23 \pm 0.03~$K, we estimated the evolutionary mass of Kn 61 by interpolation among the \citet{2006A&A...454..845M} post-AGB evolutionary tracks. The statistical uncertainty was estimated via Monte-Carlo propagation of the effective temperature and luminosity errors, yielding $m_{\mathrm{evol}}=0.575_{-0.015}^{+0.017}~\mathrm{M}_{\odot}$. The corresponding $1-\sigma$ statistical uncertainty was taken as the mean of the upper and lower errors (0.016 $\mathrm{M}_{\odot}$). A systematic uncertainty of 0.01 $\mathrm{M}_{\odot}$ was adopted to account for finite grid resolution and interpolation method dependence. The total uncertainty was obtained by adding both contributions in quadrature, resulting in $m_{\mathrm{evol}}=0.575(19)~\mathrm{M}_{\odot}$. The asteroseismic and evolutionary masses are consistent within the uncertainties ($1.2\sigma)$.

\section{Recurring brightening events} \label{sec:outbursts}

\begin{figure*}
    \centering
    \includegraphics[width=0.8\textwidth]{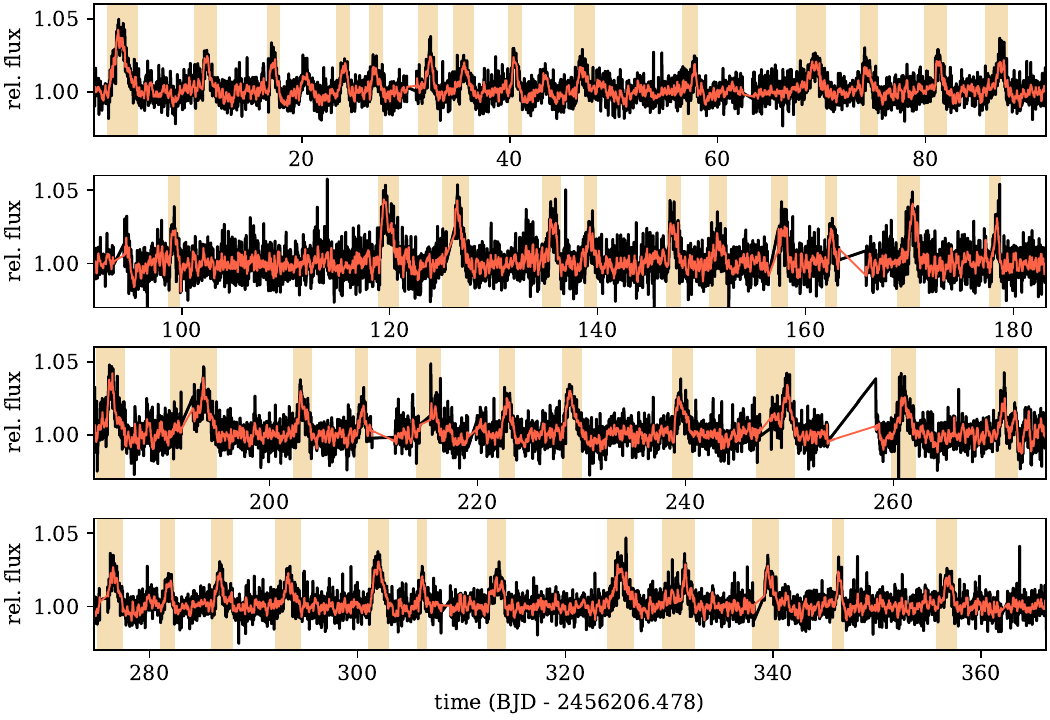}
    \caption{Long-cadence \textit{Kepler} light curves from Quarters 10-13 with 49 significant brightness increases highlighted. The detrended long-cadence data are shown in black, the smoothed version used to detect brightening events is red.}
    \label{fig:outbursts}
\end{figure*}

Besides the pulsational variability, the \textit{Kepler} light curves of Kn 61 reveal occasional increases in brightness that were noted by \citet{2015MNRAS.448.3587D}. We inspect the \textit{Kepler} pixel data to confirm that this signal is only present in the pixels surrounding the target, which is not crowded by any brighter contaminant stars. These events appear similar in their amplitudes and timescales to the pulsational outbursts observed from cool hydrogen-atmosphere pulsating white dwarfs \citep[DAVs;][]{2015ApJ...809...14B,2016ApJ...829...82B}. We modify the analysis pipeline for characterizing DAV outbursts in the \textit{Kepler} and \textit{K2} data (Bell et al.\ in prep) to analyze the brightening events of Kn 61.

With durations upwards of a day, the long-cadence (30-minute) \textit{Kepler} light curves are sufficient for detecting the brightenings and have less scatter than the short-cadence data. To remove long timescale trends, we flatten the light curve with a 5-day sliding window using the Tukey biweight algorithm implemented in {\tt WOTAN} \citep{Wotan}. We use the same approach with an 8-hour window to obtain a smoothed version of the light curve. The detrended light curve is displayed in Figure~\ref{fig:outbursts}, with the smoothed version shown in red. We flag any instance where five consecutive points in the smoothed data exceed three times the standard deviation of the smoothed light curve as a significant brightening event. Because the brightening events inflate the standard deviation of the smoothed curve, we repeat the process with the detected events masked until the number of detected outbursts converges, resulting in the 49 occurrences highlighted in Figure~\ref{fig:outbursts}. 

\begin{figure*}
    \centering
    \includegraphics[width=0.7\textwidth]{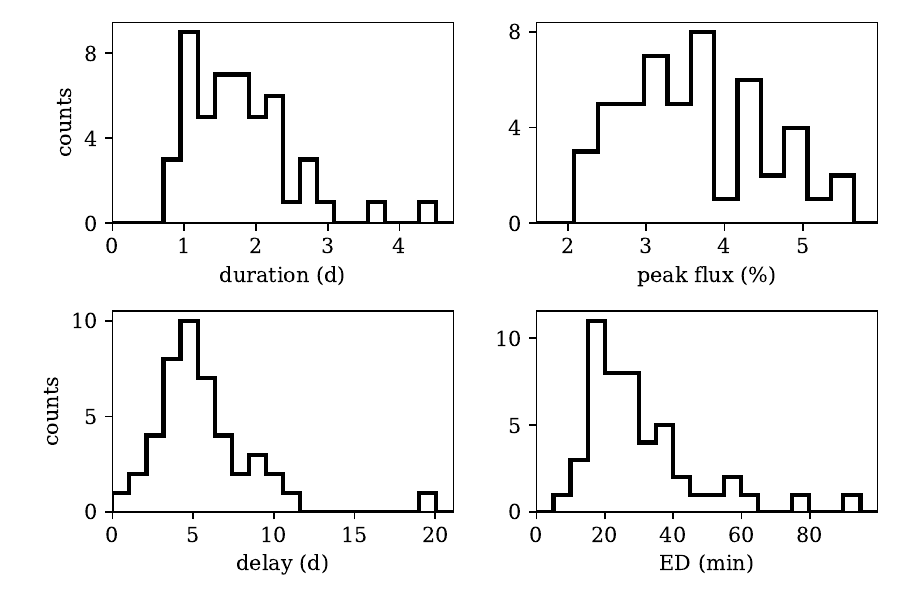}
    \caption{Histograms displaying properties of brightening events measured from the \textit{Kepler} light curve of Kn 61. Distributions are shown for event duration (in days), peak relative flux enhancement from the smoothed light curve (in percent), delay time since the previous event (days), and the equivalent duration (ED, in minutes; see text for description).}
    \label{fig:histograms}
\end{figure*}

We record the start and end times of each event when the smoothed curve crosses its median level, the peak relative flux value of the smoothed curve, the time since the previous detected event (if there are no intervening gaps in the data). Due to limitations from the signal-to-noise of the data, our detections are likely incomplete for small events. We also numerically integrate the excess relative flux during each event as the equivalent duration \citep[ED;][]{Gershberg1972}. ED has units of time representing how long the star would have to shine in quiescence to emit the same energy in the \textit{Kepler} bandpass as the excess energy from the brightening event. Figure~\ref{fig:histograms} shows histograms displaying the distributions of event durations, peak flux enhancements, delay times, and EDs. Events significant enough to be detected in the data set tend to last 1-3 days, increasing the flux in the \textit{Kepler} bandpass by 2-5\%. The distribution of delay times between events is strongly peaked at 5 days and the median ED is 26 minutes.  Overall, these observed characteristics are similar to those of pulsational outbursts of DAVs \citep{2015ApJ...809...14B,2016ApJ...829...82B}.

We can estimate the energy powering each of these events if we assume that the increase in brightness is due to an increase in the effective temperature of the target, by considering how the brightness increase in the \textit{Kepler} bandpass relates to the overall increase in bolometric luminosity. To compute how a fractional increase in \textit{Kepler} flux relates the overall luminosity increase, we compute model spectra for PG 1159 stars with the surface abundances of Kn 61 \citep{2023MNRAS.521..668B} and effective temperatures surrounding that of Kn 61 (150-200 kK in 10 kK step) with TMAW tool \citep{2018MNRAS.475.3896R}. We then multiply the model spectra by the \textit{Kepler} instrument response function\footnote{\url{https://keplergo.github.io/KeplerScienceWebsite/the-kepler-space-telescope.html\#instrument-response}} and integrate to estimate the \textit{Kepler} flux relative to the bolometric luminosity. Using a linear approximation, we find that a small fractional increase in flux in the \textit{Kepler} band corresponds to a 4.2 times larger fractional increase in luminosity overall. This is greater than one because the spectrum of Kn 61 peaks in the far UV and \textit{Kepler} observes at optical wavelengths. The energy of each event can be estimated as $4.2 L_\star\times{\rm ED}$. Using $\log{L_{\star}}=3{.}24\pm0.17\ L_{\odot}$, we estimate the median energy of a brightening event to be $\approx 4\times10^{40}$\,erg. 
Since the mechanism behind these events is not established to relate to an increase in effective temperature of the target, this approach to estimating their energies may not be accurate.

We note that spectroscopic outbursts have also been reported in Longmore~4 (Lo~4), another pulsating PG~1159/GW~Vir star in a planetary nebula \citep{1990AJ....100..788B, 1992A&A...259L..69W}. Although the outbursts in Lo~4 appear to recur on timescales of $\sim$100~days \citep{2014AJ....148...44B} -- longer than those observed in Kn~61 -- this estimate is based on sparse spectroscopic monitoring and carries significant uncertainty. Moreover, the Lo~4 events involve mass-loss episodes that change the stellar spectral type \citep{1992A&A...259L..69W}, and may not be physically related to the photometric outbursts reported here.

\section{Discussion and conclusions}
The discovery of multiperiodic $g$-mode pulsations in Kn 61 categorizes it as a GW Vir pulsator. This finding is particularly significant as Kn 61 is a N-rich PG 1159 star, adding to the growing evidence supporting the nitrogen dichotomy hypothesis, in which all N-rich PG 1159 stars are pulsators. Previous ground-based observations by \citet{2014AJ....148...57G} failed to detect these pulsations, likely because of insufficient photometric precision and short duration of observations, for which the \textit{Kepler} mission is superior in providing data for asteroseismic studies. 

Our frequency analysis reveals a sparse spectrum with four significant peaks detected down to 1.86 mmag amplitudes, which is rather typical for GW Vir stars. The identification of a possible rotational splitting indicates a potential rotation period of about 2.22 days. Such a rotation period would be rather long for a star of its type, however not impossible nor is it completely out of line with measurements of more evolved white dwarfs from asteroseismology \citep{2017ApJS..232...23H}. However, with such a small number of modes detected, the possibility of misidentification is relatively high and it cannot be ruled out with the present dataset.

The most critical finding from an asteroseismic perspective is the detection of a mean period spacing of $\Delta\Pi=21.526(6)$~s for what we identify as a sequence of $\ell=1$ modes. This regular spacing, as predicted by the asymptotic theory for chemically homogeneous stars, allows us to probe the interior structure of the star. Because PG 1159 stars are chemically stratified we can expect departures from the uniform pattern due to mode trapping. In the case of Kn 61 we identified a sequence of three modes that fit a linear progression. 
The apparent absence of many of the expected modes in the sequence could be attributed to, among others, strong mode trapping for those missing modes, those modes not being excited in the star, or not being observed due to observational biases.

The detected asymptotic period spacing is crucial for estimating the asteroseismic mass of the star. Using the relation between $\Pi_0$ and stellar mass for GW Vir stars provided by \citet{1994ApJ...427..415K}, we derived an asteroseismic mass of $m_{\mathrm{astero}}=0.551(6)~\mathrm{M}_{\odot}$. This value shows a reasonable agreement with the evolutionary mass of $m_{\mathrm{evol}}=0.575(19)~\mathrm{M}_{\odot}$ derived from the evolutionary tracks, which was calculated by interpolating the star's position on the H-R diagram against the evolutionary tracks of \citet{2006A&A...454..845M}. The consistency between these two mass determinations reinforces the validity of our asteroseismic analysis and the underlying stellar evolution models.

Recently, \citet{2024A&A...691A.194C} conducted a comprehensive comparison of WD and pre-WD mass determination methods, highlighting the strengths and limitations of both spectroscopic and asteroseismic techniques. Their study shows that while spectroscopic masses can be affected by model atmosphere uncertainties, asteroseismic ones can be more robust when multiple consecutive modes are detected. In our case, despite the limited number of identified periods, the consistency of the derived spacing with theoretical expectations supports the asteroseismic estimate of the mass of Kn 61. Our result agrees well with the trends shown in \citet{2024A&A...691A.194C}, highlighting that even relatively sparse pulsation spectra are useful for constraining fundamental stellar parameters in GW Vir stars.

In addition to the pulsations, the \textit{Kepler} light curve exhibits sporadic brightening events that typically last 1-3 days, increasing the relative flux in the light curve upwards of 5\%, with an average recurrence time of 5 days. Our estimate of the typical energy of these events is $10^{40}$\,erg, far exceeding the energies of pulsational outbursts in DAVs of $\sim10^{34}$\,erg \citep{2016ApJ...829...82B}, due mostly to the greater luminosity of PG 1159 stars. Outbursts in DAVs are thought to get their energy from pulsations \citep{2018ApJ...863...82L}, the pulsation in these stars are observed to be affected by the outbursts \citep{2015ApJ...810L...5H}. It is not clear that the brightening events in Kn 61 are related to DAV outbursts, as interpreting them this way implies that they are a million times more energetic yet they do not appear to affect the stability of the observed pulsations. Still, it is intriguing that the only PG 1159 star observed to show this behavior is a pulsator. Our energy estimates rely on the assumption that the brightness increase can be fully attributed to an increase in effective temperature of the star, which may not be valid. More theoretical work is needed to determine the nature of these events.

The discovery adds a new, multiperiodic pulsator to the group of GW Vir stars, making Kn 61 the 12th pulsating PG~1159 star with an asteroseismic mass determination, and the 15th such GW Vir star (including pulsating [WC] stars). As a confirmed N-rich pulsator, Kn 61 strengthens the proposed connection between nitrogen abundance and pulsations in PG 1159 stars. 
The detection of a clear period spacing establishes Kn 61 as a valuable target for detailed future asteroseismic modeling.

%% Please use the acknowledgment and contribution environments. This will 
%% be anonomyized when the "anonymous" style option is used. 

\begin{acknowledgments}

We thank the anonymous referee for constructive comments that improved the manuscript.
PS thanks Thomas Rauch for his assistance in retrieving the TMAW models.
This research was supported in part by the Polish National Center for Science (NCN) through grants 2015/18/A/ST9/00578 and 2021/43/B/ST9/02972.  PS acknowledges support from the Agencia Estatal de Investigaci\'on del Ministerio de Ciencia, Innovaci\'on y Universidades (MCIU/AEI) under grant ``Revolucionando el conocimiento de la evoluci\'on de estrellas poco masivas'' and the European Union NextGenerationEU/PRTR with reference CNS2023-143910 (DOI:10.13039/501100011033).
KJB is supported by the National Science Foundation under grant No.\ AST-2406917, and by NASA through grant No.\ 80NSSC25K0122 of the TESS Cycle 7 General Investigator Program. JJH is supported by NASA under Grant No.\ 80NSSC25K7902. 
This paper includes data collected by the \textit{Kepler} mission. Funding for the \textit{Kepler} mission is provided by the NASA Science Mission Directorate. STScI is operated by the Association of Universities for Research in Astronomy, Inc., under NASA contract NAS 5–26555.
The TMAW tool (\href{http://astro.uni-tuebingen.de/~TMAW}{http://astro.uni-tuebingen.de/~TMAW}) used for this paper was constructed as part of the activities of the German Astrophysical Virtual Observatory.

\end{acknowledgments}

\begin{contribution}
JJH suggested the project and validated the analysis. PS performed the analysis, prepared the figures and wrote the manuscript. GH supervised and validated the work. KJB contributed the section on brightening events. All co-authors discussed the results and provided comments and corrections on the manuscript.

\end{contribution}

\facility{Kepler}

% \software{astropy \citep{2013A&A...558A..33A,2018AJ....156..123A,2022ApJ...935..167A},  
%           Cloudy \citep{2013RMxAA..49..137F}, 
%           Source Extractor \citep{1996A&AS..117..393B}
%           }
\software{\texttt{matplotlib} \citep{Hunter:2007}, 
          \texttt{numpy} \citep{2020Natur.585..357H},
          \texttt{scipy} \citep{2020SciPy-NMeth},
          \texttt{lightkurve} \citep{2021zndo...1181928B},
          \texttt{Period04} \citep{2005CoAst.146...53L}}

\bibliography{Kn61_RNAAS}{}
\bibliographystyle{aasjournalv7}

\end{document}